\documentclass[prb,preprintnumbers,amsmath,amssymb,floatfix]{revtex4}

\usepackage{graphicx}
\usepackage{subfigure}
\usepackage{dcolumn}

\begin{document}

\title{Quantum theory of two-photon interference}
\author{Xiang-Yao Wu$^{a}$ \footnote{E-mail: wuxy2066@163.com}, Bo-Jun Zhang$^{a}$,
Xiao-Jing Liu$^{a}$, Hong Li$^{a}$  \\ Si-Qi Zhang$^{a}$, Jing
Wang$^{a}$, Yi-Heng Wu$^{b}$ and Jing-Wu Li$^{c}$}
\affiliation{a.Institute of Physics, Jilin Normal
University, Siping 136000, China \\
b. College of physics, Jilin University, Changchun 130000, China
\\
c. Institute of Physics, Xuzhou Normal University, Xuzhou 221000,
China }

\begin{abstract}
In this paper, we study two-photon interference with the approach
of photon quantum theory, with specific attention to the
two-photon interference experiment carried out by Milena D'Angelo
et al. $(Phys. Rev. Lett 87:013602, 2001)$. We find the
theoretical result is accordance with experiment data. 

\end{abstract}
\maketitle

{\bf 1. Introduction} \vskip 8pt

Nonclassical interference is one of the most remarkable phenomena
in quantum optics. In particular, it can be observed in
experiments with spontaneous parametric down conversion (SPDC)
[1], a nonlinear optical process in which high-energy pump photons
are converted into pairs of low-energy photons (usually called
signal and idler) inside a crystal with quadratic nonlinearity. It
has been shown in many experiments that the quantum state of the
signal-idler photon pair is entangled [2]. Many experiments have
made use of SPDC to demonstrate fascinating topics in quantum
optics, such as the test of Bell¡¯s inequalities, quantum
communication, quantum teleportation, etc. [3], and its possible
applications include quantum communication, computation, and
cryptography [4]. All these experiments basically belong to the
same category: quantum interference. Two-photon interference is
one of the pure quantum phenomena attributed to quantum
correlations. In experiments on two-photon interference, each
photon pair behaves like a quantum object called a "biphoton",
whose effective energy (or frequency) is twice that of the
original photons, and the interference fringe of the photon pair
has half the period of a one-photon interference fringe.

Two-photon interference is a powerful tool to study the
fundamental problems of quantum theory. For example, the
Einstein-Podolsky-Rosen problem [5] is believed to be resolvable
by testing Bell¡¯s inequality [6] and the Greenberger-
Horne-Zeilinger theorem [7] in two-photon or multiphoton
interference experiments. Two-photon interferometry also has broad
applications in practical areas such as quantum cryptography [8],
metrology [9], potentially in quantum computing [10], precision
metrology, information processing and imaging Coincidence imaging,
or ghost imaging [11, 12].

Recently it has been argued that classically correlated light
might mimic some features of the entangled photon pairs in
coincidence imaging setups. Notice that the possibility of
simulating the two-photon imaging features of entangled states
with classical sources was not ruled out by the authors of the
original ghost imaging experiment [13]. Both the theoretical work
of Abourraddy et.al. [14] and the experimental investigation of
Bennink et.al. [15] stimulated a very interesting debate about the
role of entanglement in two-photon coincidence imaging [16]. In
this work, we study the two-photon interference with the approach
of relativistic quantum theory of photon. In the viewpoint of
quantum theory, the light has the nature of wave, and it is
described by wave function $\vec{\psi}(\vec{r},t)$ for the photon
of spin $1$ . The absolute square $|\vec{\psi}(\vec{r},t)|^{2}$
can be explained as the photon's probability density at the
definite position. For light interference and diffraction, the
interference and diffraction intensity $I$ is directly
proportional to $|\vec{\psi}(\vec{r},t)|^{2}$ distributing on
display screen, and the light wave functions can be divided into
three areas. The first area is the incident area, where the photon
wave function is a plane wave. The second area is the slit area,
where the light wave function can be calculated by quantum wave
equation of photon. The third area is the diffraction area, where
the light wave function can be calculated by the Kirchhoff's law.
For double-slit interference, we can obtain the total diffraction
wave function by superposition the diffraction wave function of
every slit. For two-photon double-slit interference, we calculate
the total interference wave function
$\vec{\psi}_{s}(\vec{r},t)=c_1\vec{\psi}_{1}(\vec{r},t)+c_2\vec{\psi}_{2}(\vec{r},t)$
and
$\vec{\psi}_{i}(\vec{r},t)=c_3\vec{\phi}_{1}(\vec{r},t)+c_4\vec{\phi}_{2}(\vec{r},t)$
for the the signal and idler photon, and the detectors $D_1$ and
$D_2$ measure the interference intensities are directly
proportional to $|\vec{\psi}_{s}(\vec{r},t)|^2$ and
$|\vec{\psi}_{i}(\vec{r},t)|^2$ for the signal and idler photon,
respectively. The intensity of coincidence measurement is directly
proportional to
$|\vec{\psi}_{s}(\vec{r},t)\cdot\vec{\psi}_{i}(\vec{r},t)|^2$ for
two-photon double-slit interference. In the following, we shall
calculate these wave functions, and compare the calculation result
with the experiment.

\vskip 8pt

\setlength{\unitlength}{0.1in}
 \begin{center}
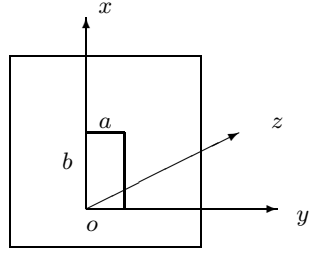
\begin{figure}
\begin{picture}(100,10)
 \put(30,4){\vector(1,0){10}}
 \put(30,4){\vector(0,1){10}}
 \put(30,4){\vector(2,1){8}}
 \put(26,2){\line(1,0){10}}
 \put(26,2){\line(0,1){10}}
 \put(36,2){\line(0,1){10}}
 \put(26,12){\line(1,0){10}}
 \put(30,8){\line(1,0){2}}
 \put(32,4){\line(0,1){4}}
 \put(41,3){\makebox(2,1)[l]{$y$}}
 \put(30,14){\makebox(2,1)[c]{$x$}}
 \put(30,2.6){\makebox(2,1)[l]{$o$}}
 \put(39,8){\makebox(2,1)[c]{$z$}}
 \put(28,6){\makebox(2,1)[c]{$b$}}
 \put(30,8){\makebox(2,1)[c]{$a$}}
\end{picture}
\caption{The single-slit geometry, $a$ is the width and $b$ is the
length of the slit. } \label{moment}
\end{figure}
\end{center}

{\bf 2. Quantum approach
 of photon single-slit diffraction  }
\vskip 8pt

In an infinite plane, we consider a single-slit, its width $a$ and
length $b$ are shown in FIG. 1. The $x$ axis is along the slit
length and the axis is along the slit width $a$. In the following,
we calculate the light wave function in the single-slit with
relativistic wave equation. At time $t$, we suppose that the
incident plane wave travels along the $z$ axis. It is
\begin{eqnarray}
\vec{\psi}_{0}(z,
t)&=&\vec{A}e^{\frac{i}{\hbar}(pz-Et)}\nonumber\\
&=&\sum_{j}A_{j}\cdot e^{\frac{i}{\hbar}(pz-Et)}\vec{e}_{j}\nonumber\\
&=&\sum_{j}\psi_{0j}\cdot e^{-\frac{i}{\hbar}Et}\vec{e}_{j},
\end{eqnarray}
where $\psi_{0j}=A_{j}\cdot e^{\frac{i}{\hbar}pz}$,  $j= x, y, z$
and $\vec{A}$ is a constant vector. The time-dependent
relativistic wave equation of light is [12]
\begin{equation}
i\hbar\frac{\partial}{\partial
t}\vec{\psi}(\vec{r},t)=c\hbar\nabla\times\vec{\psi}(\vec{r},t)+V\vec{\psi}(\vec{r},t),
\end{equation}
where $c$ is light velocity. From Eq. (2), we can find the light
wave function $\vec{\psi}(\vec{r},t)\rightarrow 0$  when
$V(\vec{r})\rightarrow\infty$. The potential energy of light in
the single-slit is
\begin{eqnarray}
V(x,y,z)&=&0  \hspace{0.3in}0\leq x\leq b, 0\leq y\leq a, 0\leq z\leq c'\nonumber\\
       &=&\infty \hspace{0.3in} otherwise,
\end{eqnarray}
where $c'$ is the slit thickness. We can get the time-dependent
relativistic wave equation in the slit ($V(x,y,z)=0$), it is
\begin{equation}
i\hbar\frac{\partial}{\partial
t}\vec{\psi}(\vec{r},t)=c\hbar\nabla\times\vec{\psi}(\vec{r},t),
\end{equation}
by derivation on  Eq. (4) about the time t and multiplying
$i\hbar$ both sides, we have
\begin{equation}
(i\hbar)^2\frac{\partial^2}{\partial
t^2}\vec{\psi}(\vec{r},t)=c\hbar\nabla\times
i\hbar\frac{\partial}{\partial t}\vec{\psi}(\vec{r},t),
\end{equation}
substituting Eq. (4) into (5), we have
\begin{eqnarray}
\frac{\partial^2}{\partial t^2}\vec{\psi}(\vec{r},t)&
=&-c^2[\nabla(\nabla\cdot\vec{\psi}(\vec{r},t))-\nabla^2\vec{\psi}(\vec{r},t)],
\end{eqnarray}
where the formula $\nabla\times\nabla\times
\vec{B}=\nabla(\nabla\cdot\vec{B})-\nabla^2\vec{B}$. From Ref.
[11], the photon wave function is
$\vec{\psi}(\vec{r},t)=\sqrt{\frac{\varepsilon_{0}}{2}}(\vec{E}(\vec{r},t)+i\sigma
c\vec{B}(\vec{r},t))$, we have
\begin{equation}
\nabla\cdot\vec{\psi}(\vec{r},t)=0,
\end{equation}
from Eq. (6) and (7), we have
\begin{equation}
(\frac{\partial^2}{\partial
t^2}-c^2\nabla^2)\vec{\psi}(\vec{r},t)=0.
\end{equation}

The Eq. (8) is the same as the classical wave equation of light.
Here, it is a quantum wave equation of light, since it is obtained
from the relativistic wave equation (2), and it satisfied the new
quantum boundary condition: when $\vec{\psi}(\vec{r},t)\rightarrow
0$, $V(\vec{r})\rightarrow\infty$. It is different from the
classic boundary condition.

When the photon wave function $\vec{\psi}(\vec{r},t)$ change with
determinate frequency $\omega$, the wave function of photon can be
written as
\begin{equation}
\vec{\psi}(\vec{r},t)=\vec{\psi}(\vec{r})e^{-i\omega t},
\end{equation}
substituting Eq. (9) into (8), we can get
\begin{equation}
\frac{\partial^{2}\vec{\psi}(\vec{r})}{\partial
x^{2}}+\frac{\partial^{2}\vec{\psi}(\vec{r})}{\partial
y^{2}}+\frac{\partial^{2}\vec{\psi}(\vec{r})}{\partial
z^{2}}+\frac{4\pi^{2}}{\\\lambda^{2}}\vec{\psi}(\vec{r})=0,
\end{equation}
and the wave function satisfies boundary conditions
\begin{equation}
\vec{\psi}(0,y,z)=\vec{\psi}(b,y,z)=0,
\end{equation}
\begin{equation}
\vec{\psi}(x,0,z)=\vec{\psi}(x,a,z)=0.
\end{equation}

The photon wave function $\vec{\psi}(\vec{r})$ can be wrote
\begin{eqnarray}
\vec{\psi}(\vec{r})&=&\psi_{x}(\vec{r})\vec{e}_{x}+\psi_{y}(\vec{r})\vec{e}_{y}+\psi_{z}(\vec{r})\vec{e}_{z}\nonumber\\
&=&\sum_{j=x,y,z}\psi_{j}(\vec{r})\vec{e}_{j},
\end{eqnarray}
where $j$ is $x$, $y$ or $z$. Substituting Eq. (13) into (10),
(11) and (12), we have the component equation
\begin{equation}
\frac{\partial^{2}\psi_{j}(\vec{r})}{\partial
x^{2}}+\frac{\partial^{2}\psi_{j}(\vec{r})}{\partial
y^{2}}+\frac{\partial^{2}\psi_{j}(\vec{r})}{\partial
z^{2}}+\frac{4\pi^{2}}{\\\lambda^{2}}\psi_{j}(\vec{r})=0.
\end{equation}
\begin{equation}
\psi_{j}(0,y,z)=\psi_{j}(b,y,z)=0,
\end{equation}
\begin{equation}
\psi_{j}(x,0,z)=\psi_{j}(x,a,z)=0.
\end{equation}

The partial differential equation (14) can be solved by the method
of separation of variable. By writing
\begin{equation}
\psi_{j}(x,y,z)=X_{j}(x)Y_{j}(y)Z_{j}(z).
\end{equation}

From Eq. (14), (15), (16) and (17), we can get the general
solution of Eq. (14)
\begin{equation}
\psi_{j}(x,y,z)=\sum_{mn}D_{mnj}\sin{\frac{n\pi
x}{b}}\sin{\frac{m\pi
y}{a}}e^{i\sqrt{\frac{4\pi^{2}}{\lambda^{2}}-\frac{n^{2}\pi^{2}}{b^{2}}-\frac{m^{2}\pi^{2}}{a^{2}}}z},
\end{equation}
since the wave functions are continuous at $z=0$, we have
\begin{equation}
\vec{\psi}_{0}(x,y,z;t)\mid_{z=0}=\vec{\psi}(x,y,z;t)\mid_{z=0},
\end{equation}
or, equivalently,
\begin{eqnarray}
\psi_{0j}(x,y,z)\mid_{z=0}&=&\psi_{j}(x,y,z)\mid_{z=0}.\hspace{0.3in}(j=x,y,z)
\end{eqnarray}

From Eq. (1), (18) and (20), we obtain the coefficient $D_{mnj}$
by fourier transform
\begin{eqnarray}
D_{mnj}&=&\frac{4}{a
b}\int^{a}_{0}\int^{b}_{0}A_{j}\sin{\frac{n\pi
\xi}{b}}\sin{\frac{m\pi \eta}{a}}d\xi d\eta \nonumber\\
&=&\frac{16A_{j}}{mn\pi^{2}} \hspace{0.6in} m,n, odd  \nonumber\\
&=&0 \hspace{0.9in} otherwise,\hspace{0.4in}(j=x,y,z)
\end{eqnarray}
substituting Eq. (21) into (18), we have
\begin{eqnarray}
\psi_{j}(x,y,z)&=&\sum_{m,n=0}^{\infty}\frac{16A_{j}}{(2m+1)(2n+1)\pi^{2}}\sin{\frac{(2n+1)\pi
x}{b}}\sin{\frac{(2m+1)\pi y}{a}} \nonumber\\&&
e^{i\sqrt{\frac{4\pi^{2}}{\lambda^{2}}-\frac{(2n+1)^{2}\pi^{2}}{b^{2}}
-\frac{(2m+1)^{2}\pi^{2}}{a^{2}}}z}, \hspace{0.6in} (j=x,y,z)
\end{eqnarray}
substituting Eq. (22) into (9) and (13), we can obtain the photon
wave function in slit
\begin{eqnarray}
\vec{\psi}(x,y,z;t)&=&\sum_{j=x,y,z}\psi_{j}(x,y,z,t)\vec{e}_{j}\nonumber\\
&=&\sum_{j=x,y,z}\sum_{m,n=0}^{\infty}\frac{16A_{j}}{(2m+1)(2n+1)\pi^{2}}\sin{\frac{(2n+1)\pi
x}{b}}\sin{\frac{(2m+1)\pi y}{a}} \nonumber\\&&
e^{i\sqrt{\frac{4\pi^{2}}{\lambda^{2}}-\frac{(2n+1)^{2}\pi^{2}}{b^{2}}
-\frac{(2m+1)^{2}\pi^{2}}{a^{2}}}z}e^{-{i}{\omega}t}\vec{e}_{j}.
\end{eqnarray}
We can consider the case of limit, i.e., the slit length $b$ is
infinity, and the Eq. (8) and (10) become
\begin{equation}
\frac{\partial^{2}}{\partial
t^{2}}\vec{\psi}(y,z,t)-c^{2}(\frac{\partial^{2}}{\partial
y^{2}}+\frac{\partial^{2}}{\partial z^{2}})\vec{\psi}(y,z,t)=0,
\end{equation}
\begin{equation}
\frac{\partial^{2}\vec{\psi}(y,z)}{\partial
y^{2}}+\frac{\partial^{2}\vec{\psi}(y,z)}{\partial
z^{2}}+\frac{4\pi^{2}}{\lambda^{2}}\vec{\psi}(y,z)=0,
\end{equation}
we can easily obtain the light wave function in the single-slit
when $b\rightarrow \infty$
\begin{eqnarray}
\vec{\psi}(y,z;t)&=&\sum_{j=y,z}\psi_{j}(x,y,z,t)\vec{e}_{j}\nonumber\\
&=&\sum_{j=x,y,z}\sum_{m=0}^{\infty}\frac{4A_{j}}{(2m+1)\pi}\sin{\frac{(2m+1)\pi
y}{a}} \nonumber\\&& e^{i\sqrt{\frac{4\pi^{2}}{\lambda^{2}}
-\frac{(2m+1)^{2}\pi^{2}}{a^{2}}}z}e^{-{i}{\omega}t}\vec{e}_{j}.
\end{eqnarray}

 {\bf 3. The wave function of photon diffraction}
\vskip 8pt In the section 2, we have calculated the photon wave
function in slit. In the following, we will calculate diffraction
wave function. we can calculate the wave function in the
diffraction area. From the slit wave function component
$\psi_{j}(\vec{r},t)$, we can calculate its diffraction wave
function component $\Phi_{j}(\vec{r},t)$ by Kirchhoff's law. It
can be calculated by the formula[17]
\begin{equation}
\Phi_{j}(\vec{r},t)=-\frac{1}{4\pi}\int_{s_{0}}\frac{e^{ikr}}{r}\overrightarrow{n}\cdot[\bigtriangledown^{'}\psi_{j}
+(ik-\frac{1}{r})\frac{\overrightarrow{r}}{r}\psi_{j}]ds.
\end{equation}
the total diffraction wave function is
\begin{eqnarray}
\vec{\Phi}(\vec{r},t)&=&\sum_{j=x,y,z}\Phi_{j}(\vec{r},t)\vec{e}_{j},
\end{eqnarray}
the diffraction area is shown in FIG. 2, where
$k=\frac{2\pi}{\lambda}$ is wave vector, $s_{0}$ is the area of
the single-slit, $\overrightarrow{r}^{'}$ the position of a point
on the surface $(z=c')$, $P$ is an arbitrary point in the
diffraction area, and the $\overrightarrow{n}$ is a unit vector,
which is normal to the surface of the single-slit. From FIG. 2, we
have
\begin{eqnarray}
r&=&R-\frac{\overrightarrow{R}}{R}\cdot\overrightarrow{r}^{'}\nonumber\\
&\approx&R-\frac{\overrightarrow{r}}{r}\cdot\overrightarrow{r}^{'}\nonumber\\
 &=&R-\frac{\overrightarrow{k_{2}}}{k}\cdot\overrightarrow{r}^{'},
\end{eqnarray}
then,

\setlength{\unitlength}{0.1in}
\begin{figure}
\begin{picture}(100,20)
 \put(26,5){\vector(1,0){3}}
 \put(26,5){\vector(0,1){2.2}}
 \put(26,5){\vector(2,1){5}}

 \put(24,1){\line(1,0){2}}
 \put(24,1){\line(0,1){4}}
 \put(26,1){\line(0,1){4}}
 \put(24,5){\line(1,0){2}}

 \put(24,10){\line(1,0){2}}
 \put(24,10){\line(0,1){4}}
 \put(26,10){\line(0,1){4}}
 \put(24,14){\line(1,0){2}}

 \put(26,7){\line(3,1){11.5}}
 \put(26,7){\vector(3,1){5}}
 \put(26,5){\line(2,1){11.5}}
 \put(27.5,3.2){\makebox(2,1)[l]{$\overrightarrow{n}$}}
 \put(24,7){\makebox(2,1)[c]{$\overrightarrow{r}^{\prime}$}}
 \put(25,5){\makebox(2,1)[l]{$o$}}
 \put(31,5.5){\makebox(2,1)[c]{$\overrightarrow{R}$}}
 \put(30,9){\makebox(2,1)[c]{$\overrightarrow{r}$}}
 \put(38.5,10.5){\makebox(2,1)[l]{$P$}}
 \put(25,-0.5){\makebox(2,1)[l]{$c$}}
\end{picture}
\caption{The diffraction area of single-slit} \label{moment}
\end{figure}
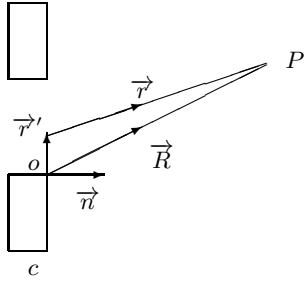
\begin{eqnarray}
\frac{e^{ikr}}{r}&=&\frac{e^{ik(R-\frac{\overrightarrow{r}}{r}\cdot\overrightarrow{r}^{'})}}
{R-\frac{\overrightarrow{r}}{r}\cdot\overrightarrow{r}^{'}}\nonumber\\
&\approx&\frac{{e^{ikR}e^{-i\overrightarrow{k_{2}}\cdot\overrightarrow{r}^{'}}}}{R}
\hspace{0.3in}(|\overrightarrow{r}^{'}|\ll R),
\end{eqnarray}
where $\vec{k_{2}}=k\frac{\vec{r}}{r}$. Substituting Eq. (22),
(29) and (30) into (27), one can obtain
\begin{eqnarray}
\Phi_{j}(\vec{r},t)&=&-\frac{e^{ikR}}{4\pi
R}e^{-{i}{\omega}t}\int_{s_{0}}e^{-i\overrightarrow{k_{2}}\cdot
\overrightarrow{r}^{'}}\sum_{m=0}^{\infty}\sum_{n=0}^{\infty}\frac{16A_{j}}{(2m+1)(2n+1)\pi^{2}}\nonumber\\&&
e^{i\sqrt{\frac{4\pi^{2}}{\lambda^{2}}-(\frac{(2n+1)\pi}{b})^{2}-(\frac{(2m+1)\pi}{a})^{2}}\cdot
c'} \sin \frac{(2n+1)\pi}{b}x^{'}\sin
\frac{(2m+1)\pi}{a}y^{'}\nonumber\\&&
[i\sqrt{\frac{4\pi^{2}}{\lambda^{2}}-(\frac{(2n+1)\pi}{b})^{2}-(\frac{(2m+1)\pi}{a})^{2}}+i\overrightarrow{n}\cdot
\overrightarrow{k_{2}}-\frac{\overrightarrow{n}\cdot
\overrightarrow{R}}{R^{2}}]dx^{'}dy^{'}.
\end{eqnarray}
Assume that the angle between $\overrightarrow{k_{2}}$ and $x$
axis ($y$ axis) is $\frac{\pi}{2}-\alpha$ ($\frac{\pi}{2}-\beta$),
and $\alpha (\beta)$ is the angle between $\overrightarrow{k_{2}}$
and the surface of $yz$ ($xz$), then we have
\begin{eqnarray}
k_{2x}=k\sin \alpha,\hspace{0.3in} k_{2y}=k\sin \beta,
\end{eqnarray}
\begin{eqnarray}
\overrightarrow{n}\cdot \overrightarrow{k_{2}}=k\cos \theta,
\end{eqnarray}
where $\theta$ is the angle between $\overrightarrow{k_{2}}$ and
$z$ axis. Substituting Eq. (32) and (33) into (31) gives
\begin{eqnarray}
\Phi_{j}(x,y,z;t)&=&-\frac{e^{ikR}}{4\pi
R}e^{-{i}{\omega}t}\sum_{m=0}^{\infty}\sum_{n=0}^{\infty}\frac{16A_{j}}{(2m+1)(2n+1)\pi^2}
e^{i\sqrt{\frac{4\pi^{2}}{\lambda^{2}}-(\frac{(2n+1)\pi}{b})^{2}-(\frac{(2m+1)\pi}{a})^{2}}\cdot
c'}\nonumber\\&&
[i\sqrt{\frac{4\pi^{2}}{\lambda^{2}}-(\frac{(2n+1)\pi}{b})^{2}-(\frac{(2m+1)\pi}{a})^{2}}+(ik-\frac{1}{R})
\sqrt{\cos^{2}\alpha-\sin^{2}\beta}]\nonumber\\&&
\int^{b}_{0}e^{-ik\sin\alpha\cdot
x^{'}}\sin\frac{(2n+1)\pi}{b}x^{'}dx^{'}\int^{a}_{0}e^{-ik\sin\beta\cdot
y^{'}} \sin \frac{(2m+1)\pi}{a}y^{'}dy^{'}.
\end{eqnarray}
Substituting Eq. (34) into (28), one can get
\begin{eqnarray}
\vec{\Phi}(x,y,z;t)&=&-\frac{e^{ikR}}{4\pi
R}e^{-{i}{\omega}t}\sum_{j=s,y,z}\sum_{m=0}^{\infty}\sum_{n=0}^{\infty}\frac{16A_{j}}{(2m+1)(2n+1)\pi^2}
e^{i\sqrt{\frac{4\pi^{2}}{\lambda^{2}}-(\frac{(2n+1)\pi}{b})^{2}-(\frac{(2m+1)\pi}{a})^{2}}\cdot
c'}\nonumber\\&&
[i\sqrt{\frac{4\pi^{2}}{\lambda^{2}}-(\frac{(2n+1)\pi}{b})^{2}-(\frac{(2m+1)\pi}{a})^{2}}+(ik-\frac{1}{R})
\sqrt{\cos^{2}\alpha-\sin^{2}\beta}]\nonumber\\&&
\int^{b}_{0}e^{-ik\sin\alpha\cdot
x^{'}}\sin\frac{(2n+1)\pi}{b}x^{'}dx^{'}\int^{a}_{0}e^{-ik\sin\beta\cdot
y^{'}} \sin \frac{(2m+1)\pi}{a}y^{'}dy^{'}\vec{e}_{j}.
\end{eqnarray}
Eq. (35) is the  total diffraction wave function in the
diffraction area. From the wave function, we can obtain the
diffraction intensity $I$ on the display screen, we have
\begin{equation}
I\propto|\vec{\Phi}(x,y,z;t)|^{2}.
\end{equation}
\vskip 8pt

{\bf 4. Double-slit diffraction wave function of photon} \vskip
8pt \setlength{\unitlength}{0.1in}

\hspace{0.2in}From Eq. (23), in the first slit, the photon wave
function $\vec{\psi}_{1}(x,y,z;t)$ is
\begin{eqnarray}
\vec{\psi}_{1}(x,y,z;t)&=&\sum_{j=x,y,z}\sum_{m,n=0}^{\infty}\frac{16A_{j}}{(2m+1)(2n+1)\pi^{2}}\sin{\frac{(2n+1)\pi
x}{b}}\sin{\frac{(2m+1)\pi y}{a}} \nonumber\\&&
e^{i\sqrt{\frac{4\pi^{2}}{\lambda^{2}}-\frac{(2n+1)^{2}\pi^{2}}{b^{2}}
-\frac{(2m+1)^{2}\pi^{2}}{a^{2}}}z}e^{-{i}{\omega}t}\vec{e}_{j}.
\end{eqnarray}

From FIG. 3, in the second slit, when we make the coordinate
translations :
\begin{eqnarray}
&&x'=x\nonumber\\
&&y'=y-(a+d)\nonumber\\
&&z'=z,
\end{eqnarray}
\begin{figure}
\begin{picture}(70,10)
 \put(26,4){\vector(1,0){18}}
 \put(26,4){\vector(0,1){10}}
 \put(26,4){\vector(2,1){15}}
 \put(22,2){\line(1,0){24}}
 \put(22,2){\line(0,1){13}}
 \put(46,2){\line(0,1){13}}
 \put(22,15){\line(1,0){24}}
 \put(26,8){\line(1,0){2}}
 \put(28,4){\line(0,1){4}}
 \put(32,4){\line(0,1){4}}
 \put(32,8){\line(1,0){2}}
 \put(34,4){\line(0,1){4}}
 \put(28,7){\line(1,0){1}}
 \put(31,7){\line(1,0){1}}
 \put(44,2.5){\makebox(2,1)[l]{$y$}}
 \put(26,13){\makebox(2,1)[c]{$x$}}
 \put(25.5,2.6){\makebox(2,1)[l]{$o$}}
 \put(41,11){\makebox(2,1)[c]{$z$}}
 \put(24,6){\makebox(2,1)[c]{$b$}}
 \put(26,8){\makebox(2,1)[c]{$a$}}
 \put(29.5,6.5){\makebox(2,1)[l]{$d$}}
 \put(27,2.6){\makebox(2,1)[l]{$1$}}
 \put(33,2.6){\makebox(2,1)[l]{$2$}}

\end{picture}
\caption{Double-slit geometry with $a$ the single slit width, $b$
the slit length and $d$ the distance between the two slit. }
\label{moment}
\end{figure}
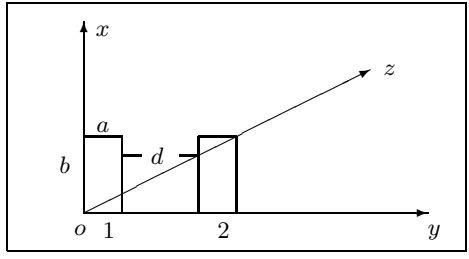
we can obtain the second slit photon wave function by the first
slit light wave function. It is
\begin{eqnarray}
\vec{\psi}_{2}(x,y,z;t)&=&\sum_{j=x,y,z}\sum_{m,n=0}^{\infty}\frac{16A_{j}}{(2m+1)(2n+1)\pi^{2}}\sin{\frac{(2n+1)\pi
x}{b}}\sin{\frac{(2m+1)\pi [y-(a+d)]}{a}} \nonumber\\&&
e^{i\sqrt{\frac{4\pi^{2}}{\lambda^{2}}-\frac{(2n+1)^{2}\pi^{2}}{b^{2}}
-\frac{(2m+1)^{2}\pi^{2}}{a^{2}}}z}e^{-{i}{\omega}t}\vec{e}_{j}.
\end{eqnarray}

With the Kirchhoff's law, similar as Eq. (35), we can get the
light diffraction wave functions of the first and second slit,
they are
\begin{eqnarray}
\vec{\Phi}_{1}(x,y,z;t)&=&-\frac{e^{ikR}}{4\pi
R}e^{-{i}{\omega}t}\sum_{j=x,y,z}\sum_{m=0}^{\infty}\sum_{n=0}^{\infty}\frac{16A_{j}}{(2m+1)(2n+1)\pi^2}
e^{i\sqrt{\frac{4\pi^{2}}{\lambda^{2}}-(\frac{(2n+1)\pi}{b})^{2}-(\frac{(2m+1)\pi}{a})^{2}}\cdot
c'}\nonumber\\&&
[i\sqrt{\frac{4\pi^{2}}{\lambda^{2}}-(\frac{(2n+1)\pi}{b})^{2}-(\frac{(2m+1)\pi}{a})^{2}}+(ik-\frac{1}{R})
\sqrt{\cos^{2}\alpha-\sin^{2}\beta}]\nonumber\\&&
\int^{b}_{0}e^{-ik\sin\alpha\cdot
x^{'}}\sin\frac{(2n+1)\pi}{b}x^{'}dx^{'}\int^{a}_{0}e^{-ik\sin\beta\cdot
y^{'}} \sin \frac{(2m+1)\pi}{a}y^{'}dy^{'}\vec{e}_{j}.
\end{eqnarray}
and
\begin{eqnarray}
\vec{\Phi}_{2}(x,y,z;t)&=&-\frac{e^{ikR}}{4\pi
R}e^{-{i}{\omega}t}\sum_{j=x,y,z}\sum_{m=0}^{\infty}\sum_{n=0}^{\infty}\frac{16A_{j}}{(2m+1)(2n+1)\pi^2}
e^{i\sqrt{\frac{4\pi^{2}}{\lambda^{2}}-(\frac{(2n+1)\pi}{b})^{2}-(\frac{(2m+1)\pi}{a})^{2}}\cdot
c'}\nonumber\\&&
[i\sqrt{\frac{4\pi^{2}}{\lambda^{2}}-(\frac{(2n+1)\pi}{b})^{2}-(\frac{(2m+1)\pi}{a})^{2}}+(ik-\frac{1}{R})
\sqrt{\cos^{2}\alpha-\sin^{2}\beta}]\nonumber\\&&
\int^{b}_{0}e^{-ik\sin\alpha\cdot
x^{'}}\sin\frac{(2n+1)\pi}{b}x^{'}dx^{'}\int^{2a+d}_{a+d}e^{-ik\sin\beta\cdot
y^{'}} \sin \frac{(2m+1)\pi}{a}y^{'}dy^{'}\vec{e}_{j}.
\end{eqnarray}
The total diffraction wave function for the double-slit is
\begin{equation}
\vec{\Phi}(x,y,z;t)=c_1\vec{\Phi}_{1}(x,y,z;t)+c_2\vec{\Phi}_{2}(x,y,z;t),
\end{equation}
where $|c_1|^2+|c_2|^2=1$. From Eq. (42), we can obtain the counts
$C$ in the detectors $D_1$ or $D_2$ is
\begin{equation}
C\propto|\vec{\Phi}(x,y,z;t)|^{2}.
\end{equation}
With the relativistic quantum theory of photon, we obtain the
relation between diffraction intensity and the slit length, slit
width, slit thickness, light wavelength and diffraction angle,
which includes all interference and diffraction information.

In the experiment of two-photon interference [18], The 458 nm line
of an Argon Ion laser is used to pump a $5mm$ $BBO$
$(\beta-BaB_2O_4)$ crystal, which is cut for degenerate collinear
type-II phase matching to produce pairs of orthogonally polarized
signal and idler photons. Each pair emerges from the crystal
collinearly, with $\omega_s \approx\omega_i\approx\omega_p/2$,
where $\omega_j$ $(j = s, i, p)$ are the frequencies of the
signal, idler and pump, respectively. The signal and idler are
interfered by the same double-slit experiment device, and the
interference-diffraction pattern of signal and idler photons are
separated by the beam splitter $PBS$ and are detected by the
photon counting detectors $D_1$ and $D_2$, respectively. The
output pulses of each detector are sent to a coincidence counting
circuit for the signal-idler joint detection. The experiment setup
is shown in FIG. 3 of Ref. [18]. Since the wavelength of signal
and idler photons are equal, and the double-slit experiment device
are same, their interference-diffraction wave function are same,
i.e., Eq. (42). The counts of photon counting detectors $D_1$ and
$D_2$ are directly proportional to $|\vec{\Phi}(x,y,z;t)|^{2}$,
and the counts of photon coincidence counting detectors $D$ is
directly proportional to
$|\vec{\Phi}(x,y,z;t)\cdot\vec{\Phi}(x,y,z;t)|^{2}$. In the
following, we shall compare the theory result with the experiment
data. \vskip 8pt

{\bf 5. Numerical result} \vskip 8pt

The double-slit interference-diffraction experiment of two-photon
had been reported by Milena D'Angelo in Ref. [18]. The experiment
parameters are: the wavelength of signal and idler photons are
$\lambda_{s}=\lambda_{i}=916nm$, the width of each slit is $a =
0.13mm$, the distance between the two slits is $d = 0.4mm$. In
theory calculation, we take the wavelength, the slit width and the
distance between the two slits are same as experiment values. In
calculation, the theory parameters are taken as: $c1=0.955$,
$c2=0.298$, $A_x=A_y=A_z=0.896$, the slit length $b=1.31\times
10^{-2}$ and the slit thickness $c=2.65\times 10^{-5}$. In solid
curve is our theoretical calculation, and the dot curve is the
experiment data [18]. From FIG. 4, we find that the theoretical
result is in accordance with the experiment data, when the
diffraction angle $\beta$ is in the range of $|\beta|\leq 2
(mrad)$. When the diffraction angle $\beta$ is in the range of
$|\beta|\geq 2 (mrad)$, the theoretical result has a small
discrepancy with the experiment data. We think the experiment data
should be measured accurately, and the theoretical calculation
should be improved furtherly.

 \vskip 8pt

{\bf 6. Conclusion} \vskip 8pt

In conclusion, we study two-photon interference with the approach
of photon quantum theory, and comparison the theoretical result
with the experimental data. We find that the calculation result is
in accordance with the experiment data, when the diffraction angle
$\beta$ is in the range of $|\beta|\leq 2 (mrad)$. When the
diffraction angle $\beta$ is in the range of $|\beta|\geq 2
(mrad)$, the theoretical result has a small discrepancy with the
experiment data. We think the experiment data should be measured
accurately, and the theoretical calculation should be improved
furtherly.

\newpage
\begin{figure}[tbp]
\begin{center}
\includegraphics[width=10 cm]{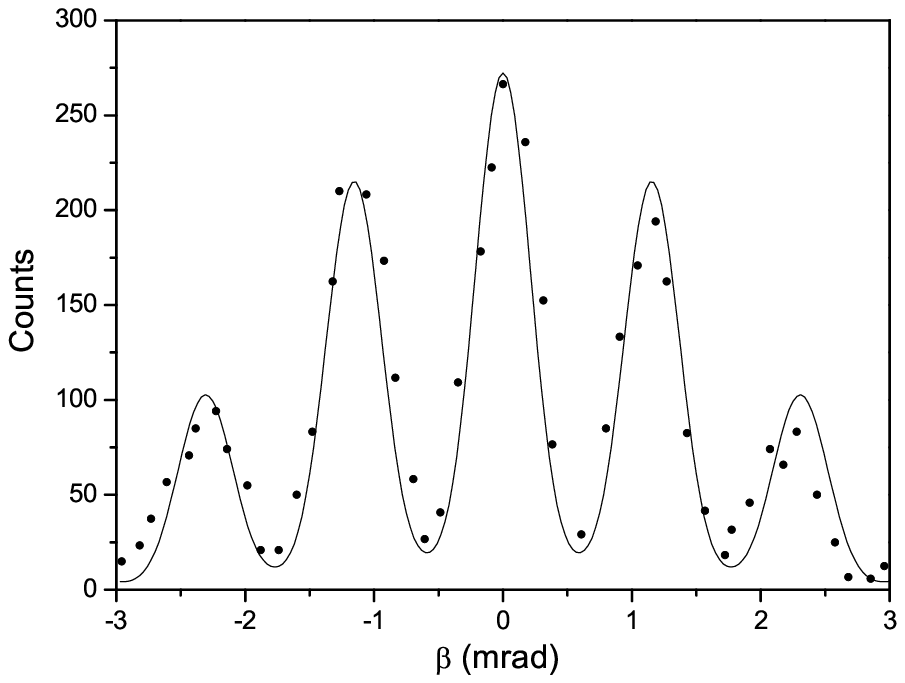}
\caption{Comparison between theoretical prediction from (43)(solid
line) and experimental data taken from [18] (circle point).}
\label{Fig1}
\end{center}
\end{figure}
\end{document}